\documentclass[conference]{IEEEtran}
\IEEEoverridecommandlockouts

\usepackage[hidelinks]{hyperref}
\usepackage{cite}
\usepackage{amsmath,amssymb,amsfonts}
\usepackage{algorithmic}
\usepackage{graphicx}
\usepackage{textcomp}
\usepackage{xcolor}
\usepackage[separate-uncertainty=true]{siunitx}
\usepackage{caption}
\usepackage{multirow}
\usepackage{bm}
\usepackage{enumitem} %
\usepackage{pgf}

\usepackage{atbegshi}
\usepackage{xsavebox}

\DeclareSIUnit\yearsold{y.o.}

\setlist{nosep, itemsep=1pt, parsep=1pt, topsep=2pt} %

\def\BibTeX{{\rm B\kern-.05em{\sc i\kern-.025em b}\kern-.08em
    T\kern-.1667em\lower.7ex\hbox{E}\kern-.125emX}}

\newif\ifshowtodos
\newcommand{\showtodos}{\showtodostrue} %
\newcommand{\hidetodos}{\showtodosfalse} %
\showtodostrue %

\hypersetup{
    pdfauthor={Paul-Otto Müller, Sven Suppelt, Mario Kupnik, and Oskar von Stryk},
    pdftitle={Jaw Tracking System: Enabling Design, Control, and Validation of Rehabilitative Jaw Exoskeletons},
    pdfsubject={Precise tracking of the jaw kinematics is crucial for diagnosing various musculoskeletal and neuromuscular diseases affecting the masticatory system and for advancing rehabilitative devices such as jaw exoskeletons, a hardly explored research field, to treat these disorders. We introduce an open-source, low-cost, precise, non-invasive, and biocompatible jaw tracking system based on optical motion capture technology to address the need for accessible and adaptable research tools. The system encompasses a complete pipeline from data acquisition, processing, and kinematic analysis to filtering, visualization, and data storage. We evaluated its performance and feasibility in experiments with four participants executing various jaw movements. The system demonstrated reliable kinematic tracking with an estimated precision of (182 ± 47) μm and (0.126 ± 0.034) °. Therefore, the open-source nature of the system and its utility comparable to commercial systems make it suitable for many research and development contexts, especially for applications such as the integration and design of jaw exoskeletons and customized diagnostic protocols. The complete system is available at GitHub with the aim of promoting innovation in temporomandibular disorders research and jaw assistive technology.},
    pdfkeywords={jaw tracking, optical motion capture, exoskeletons, rehabilitation robotics, temporomandibular disorders (TMDs), open-source, sensor systems}
}
    
\begin{document}

\hidetodos
\showtodos %

\title{An Optical Measurement System for Open-Source Tracking of Jaw Motions\\
\thanks{This work was partially supported by the German Research Foundation DFG within RTG 2761 LokoAssist (Grant no. 450821862).}
}

\author{\IEEEauthorblockN{Paul-Otto Müller\IEEEauthorrefmark{1},
Sven Suppelt\IEEEauthorrefmark{2},
Mario Kupnik\IEEEauthorrefmark{2}, and
Oskar von Stryk\IEEEauthorrefmark{1}}
\IEEEauthorblockA{\IEEEauthorrefmark{1}Simulations, Systems Optimization and Robotics Group, Technical University of Darmstadt, Darmstadt, Germany}
\IEEEauthorblockA{\IEEEauthorrefmark{2}Measurement and Sensor Technology Group, Technical University of Darmstadt, Darmstadt, Germany}
\textit{\{pmueller@sim., sven.suppelt@, mario.kupnik@, stryk@sim.\}tu-darmstadt.de}}

\makeatletter
\newcommand\notsotiny{\@setfontsize\notsotiny{6.6}{7.7}}
\makeatother

\xsavebox{PageTopWatermark}{%
    \begin{minipage}{\paperwidth}
        \notsotiny
        \centering
        Preprint of the paper which appeared in: 2025 IEEE SENSORS. DOI: 10.1109/SENSORS59705.2025.11330651
    \end{minipage}
}

\xsavebox{PageBottomWatermark}{%
    \begin{minipage}{\paperwidth}
        \notsotiny
        \centering
        \parbox{18cm}{\centering
        \textcopyright 2025 IEEE. Personal use of this material is permitted. Permission from IEEE must be obtained for all other uses, in any current or future media, including reprinting/republishing
        this material for advertising or promotional purposes, creating new collective works, for resale or redistribution to servers or lists, or reuse of any copyrighted component of this work in other works.}
    \end{minipage}
}

\AtBeginShipout{
    \AtBeginShipoutUpperLeft{\raisebox{-0.8cm}{\xusebox{PageTopWatermark}}}
}

\AtBeginShipout{
    \AtBeginShipoutUpperLeft{\raisebox{-27.3cm}{\xusebox{PageBottomWatermark}}}
}

\maketitle

\begin{abstract}

Precise tracking of the jaw kinematics is crucial for diagnosing various musculoskeletal and neuromuscular diseases affecting the masticatory system and for advancing rehabilitative devices such as jaw exoskeletons, a hardly explored research field, to treat these disorders.
We introduce an open-source, low-cost, precise, non-invasive, and biocompatible jaw tracking system based on optical motion capture technology to address the need for accessible and adaptable research tools.
The system encompasses a complete pipeline from data acquisition, processing, and kinematic analysis to filtering, visualization, and data storage.
We evaluated its performance and feasibility in experiments with four participants executing various jaw movements.
The system demonstrated reliable kinematic tracking with an estimated precision of $(182 \pm 47)\,\mu \mathrm{m}$ and $(0.126 \pm 0.034)^{\,\circ}$.
Therefore, the open-source nature of the system and its utility comparable to commercial systems make it suitable for many research and development contexts, especially for applications such as the integration and design of jaw exoskeletons and customized diagnostic protocols.
The complete system is available at \textit{GitHub} with the aim of promoting innovation in temporomandibular disorders research and jaw assistive technology.

\end{abstract}

\begin{IEEEkeywords}
    jaw tracking, optical motion capture, exoskeletons, rehabilitation robotics, temporomandibular disorders (TMDs), open-source, sensor systems
\end{IEEEkeywords}

\section{Introduction}

\begin{figure}[tbp]
    \centerline{\includegraphics[width=.95\columnwidth,scale=1.0]{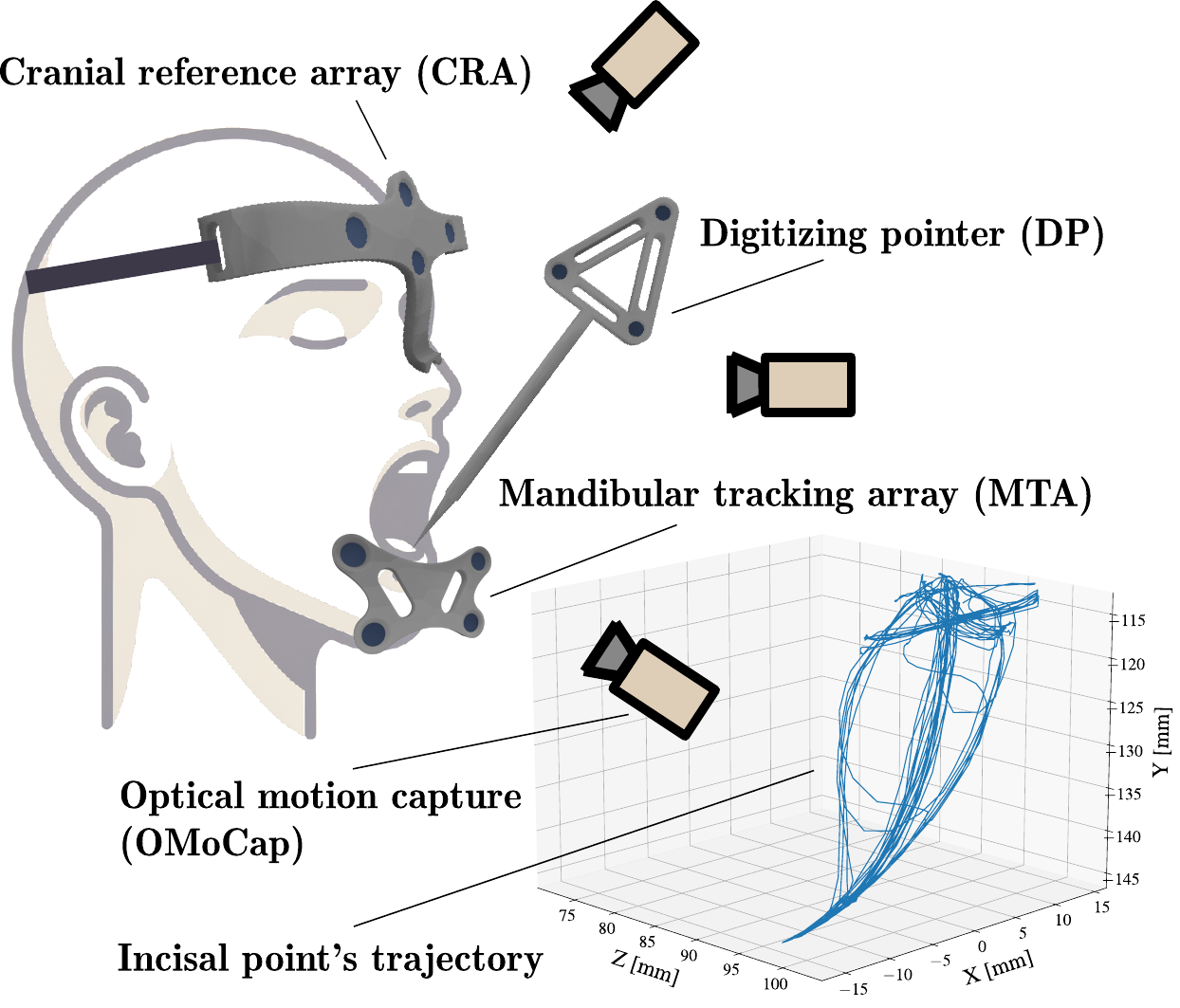}}
    \caption{The jaw tracking system consists of a mandibular tracking array attached to the lower teeth with dental glue, a cranial reference array on the forehead, and a pointer for anatomical landmark digitization. The optical motion capture system tracks the \num{3}D positions of all markers. An exemplary trajectory of the lower incisal point is shown on the lower right. The image of the head was generated with OpenAI's \textit{DALL-E} \num{3}.}
    \label{fig:jaw-tracking-setup}
\end{figure}

\begin{figure*}[tbp]
    \centerline{\includegraphics[width=.9\textwidth,scale=1.0]{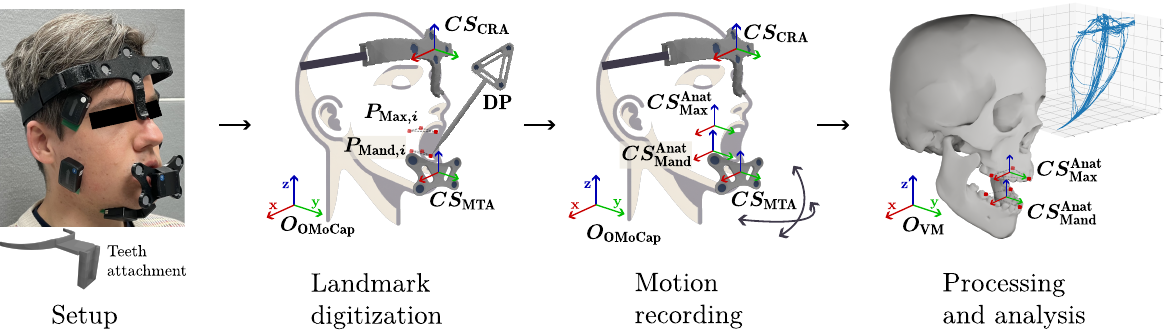}}
    \caption{Illustration of the jaw tracking system's data acquisition and processing pipeline. The setup begins with the attachment of the MTA to the lower teeth and the CRA to the forehead. The system utilizes an OMoCap system to record the $3$D positions of reflective markers attached to these arrays. The DP is used to register anatomical landmarks on the teeth, defining local coordinate systems for the mandible and maxilla. During operation, the patient performs various jaw movements, and the system records the marker trajectories. The raw data is processed to calculate relative mandibular motion, isolate jaw kinematics, and register the motion to a virtual jaw model. The final output includes filtered kinematic data, which can be visualized and stored for further analysis.}
    \label{fig:jaw-tracking-pipeline}
\end{figure*}

The complexity of the masticatory system, particularly the temporomandibular joints, responsible for essential functions like chewing and speaking, makes it susceptible to various musculoskeletal and neuromuscular conditions termed temporomandibular disorders (TMDs) \cite{Ahmad2016, Gremillion2018}. TMD symptoms include impaired masticatory function, headaches, and jaw pain, significantly deteriorating quality of life \cite{Durham2015, Tran2022, Suppelt2025}. While noninvasive treatments like physiotherapy represent the first-line approach for symptom alleviation, severe cases may require surgical intervention \cite{Shimada2019}.

Active jaw exoskeletons present a promising yet underexplored rehabilitation approach, potentially assisting therapy by providing support and collecting relevant data \cite{Mueller2025b}. Developing and evaluating sensors and control algorithms for such systems requires precise, fast, non-invasive jaw tracking systems that provide necessary simulation data or ground truth for calibrating internal sensors and actuators. Moreover, jaw kinematics serve as important indicators of rehabilitation progress and proper masticatory system function \cite{Scolaro2022, Nota2024}.

Various jaw tracking systems have emerged over the years, including commercial and non-commercial electromagnetic (\textit{AG500}, Carstens Medizinelectronik)  \cite{Morikawa2022, Jucevicius2022}, ultrasonic (\textit{WinJMA}, Zebris Medical GmbH), electromechanical (\textit{CADIAX 4}, Gamma), and optical systems (\textit{Modjaw}, ModJaw; \textit{Cyclops}, Itaka WayMed) \cite{Farronato2021, Nagy2023, Grande2024}. Each technology presents distinct advantages and limitations: electromagnetic approaches suffer from magnetic interference, ultrasonic systems face environmental noise susceptibility, and electromechanical systems have a limited range of motion---complications that become problematic when integrated with active jaw exoskeletons. Optical systems avoid these limitations while providing high accuracy and resolution, though they may require more complex setups and suffer from occlusions or reflections. All systems require the calibration and attachment of a tracker to the lower jaw.

Since existing optical jaw tracking systems are often expensive with limited flexibility and data availability, or require laborious dental casts and $3$D scans, simpler, accessible systems better suit large patient cohort evaluations and rapid jaw exoskeleton prototyping. Therefore, we present a $3$D jaw tracking system based on optical motion capture technology that is customizable, low-cost, biocompatible, and open-source, facilitating jaw exoskeleton development and TMD rehabilitation status evaluation in clinical practice (\autoref{fig:jaw-tracking-setup}). Furthermore, we describe a comprehensive processing pipeline for data acquisition, processing, analysis, visualization, and storage, and evaluate the system in experiments with four participants performing various jaw movements. 
The complete system is available at \textit{GitHub} \cite{Mueller2025c}.

\section{Jaw Tracking System}
\label{sec:jaw-tracking-system}

Our $3$D jaw tracking system enables precise mandibular kinematic analysis while minimizing soft-tissue artifacts through four key components: (1) The mandibular tracking array (MTA), a reflective marker-equipped mouthpiece rigidly attached to the lower teeth via custom, single-use adapters and temporary dental adhesive, printable with biocompatible materials (e.g. \textit{IBT Resin}, medical-grade \textit{PETG}); (2) The cranial reference array (CRA), a lightweight, marker-equipped headpiece positioned on the forehead to track head movements, enabling their subtraction from MTA data to isolate relative jaw motion; (3) The digitizing pointer (DP), a marker-equipped pointer with known geometry and pointed tip for digitizing anatomical tooth landmarks to register kinematic data to virtual jaw models; (4) An optical motion capture (OMoCap) System tracking $3$D positions of all markers, with any system providing synchronized $3$D marker coordinates compatible with the processing pipeline being suitable.

\subsection{Calibration and Data Acquisition Procedure}
The tracking procedure involves patient setup, landmark digitization, and motion recording (\autoref{fig:jaw-tracking-pipeline}):
\begin{enumerate}[left=0pt]
    \item \textbf{Setup:} The MTA is adhered to the lower teeth, and the CRA is secured on the forehead.
    \item \textbf{Landmark digitization:} With the OMoCap system active, the DP tip is used to record the \num{3}D positions of six anatomical landmarks: three on the mandibular dentition $\bm{P}_{\text{Mand},i}$ and three on the maxillary dentition $\bm{P}_{\text{Max},i}$, $i=1,2,3$. These points define anatomical coordinate systems for the mandible $CS_{\text{Mand}}^{\text{Anat}}$ and maxilla $CS_{\text{Max}}^{\text{Anat}}$.
    \item \textbf{Motion recording:} The patient performs various jaw motions. The OMoCap system records the \num{3}D trajectories of markers on the MTA and CRA.
\end{enumerate}
The rigid MTA-teeth connection ensures arbitrary MTA placement since its relative transformation to $CS_{\text{Mand}}^{\text{Anat}}$ remains fixed and is determined during calibration.

\subsection{Data Processing and Kinematic Analysis}
\label{subsec:jaw-tracking-pipeline}

Raw marker data is processed to derive relative mandibular kinematics, with $O_{\text{OMoCap}}$ denoting the global OMoCap coordinate system and all transformations represented as homogeneous matrices $\bm{T}\in\mathbb{R}^{4\times4}$ (\autoref{fig:jaw-tracking-pipeline}):

\begin{enumerate}[left=0pt]
    \item \textbf{Local coordinate systems:}
        Local coordinate systems are defined for the MTA ($CS_{\text{MTA}}$) and CRA ($CS_{\text{CRA}}$) based on their respective marker configurations. The anatomical systems $CS_{\text{Mand}}^{\text{Anat}}$ and $CS_{\text{Max}}^{\text{Anat}}$ are defined by the digitized landmarks $\bm{P}_{\text{Mand},i}$ and $\bm{P}_{\text{Max},i}$.

    \item \textbf{Static transformation:}
        From the local coordinate systems, the constant transformation $ {^{CS_{\text{MTA}}}}\bm{T}_{CS_{\text{Mand}}^{\text{Anat}}} $ is determined. 

    \item \textbf{Dynamic tracking (at time $t$):}
        The OMoCap system provides the poses $ {^{O_{\text{OMoCap}}}}\bm{T}_{CS_{\text{MTA}}}(t) $ and $ {^{O_{\text{OMoCap}}}}\bm{T}_{CS_{\text{CRA}}}(t) $ of the MTA and CRA in the global frame, respectively.

    \item \textbf{Relative mandibular motion calculations:}
        The primary goal is to determine the motion of the mandible relative to the global frame of the virtual model $^{O_{\text{VM}}}\bm{T}_{CS_{\text{Mand}}^{\text{Anat}}} (t)$.
        First, we find the pose of the anatomical mandible frame in the CRA frame, 
        \begin{align*}
            &{^{CS_{\text{CRA}}}}\bm{T}_{CS_{\text{Mand}}^{\text{Anat}}} (t) = \\
            &({^{O_{\text{OMoCap}}}}\bm{T}_{CS_{\text{CRA}}}(t))^{-1} \cdot {^{O_{\text{OMoCap}}}}\bm{T}_{CS_{\text{MTA}}}(t) \cdot {^{CS_{\text{MTA}}}}\bm{T}_{CS_{\text{Mand}}^{\text{Anat}}} .
        \end{align*}
        Then, we use the Kabsch algorithm \cite{Kabsch1976} and $\bm{P}_{\text{Max},i}$, defined in $O_{\text{VM}}$ and $CS_{\text{CRA}}$, to find the optimal rotation and translation between the two sets of points, leading to the transformation $ {^{O_{\text{VM}}}}\bm{T}_{CS_{\text{CRA}}} $.  
        Finally, we get
        \begin{align*}
            ^{O_{\text{VM}}}\bm{T}_{CS_{\text{Mand}}^{\text{Anat}}} (t) &= {^{O_{\text{VM}}}}\bm{T}_{CS_{\text{CRA}}} \cdot {^{CS_{\text{CRA}}}}\bm{T}_{CS_{\text{Mand}}^{\text{Anat}}} (t) .
        \end{align*}

    \item \textbf{Filtering, visualization, and storage:}
        $^{O_{\text{VM}}}\bm{T}_{CS_{\text{Mand}}^{\text{Anat}}}(t)$ undergoes bidirectional Savitzky-Golay filtering for noise reduction, two- or three-dimensional visualization, and \textit{HDF5} format storage for further analysis.
\end{enumerate}

\begin{figure}[tbp]
    \centerline{\includegraphics[width=.9\columnwidth]{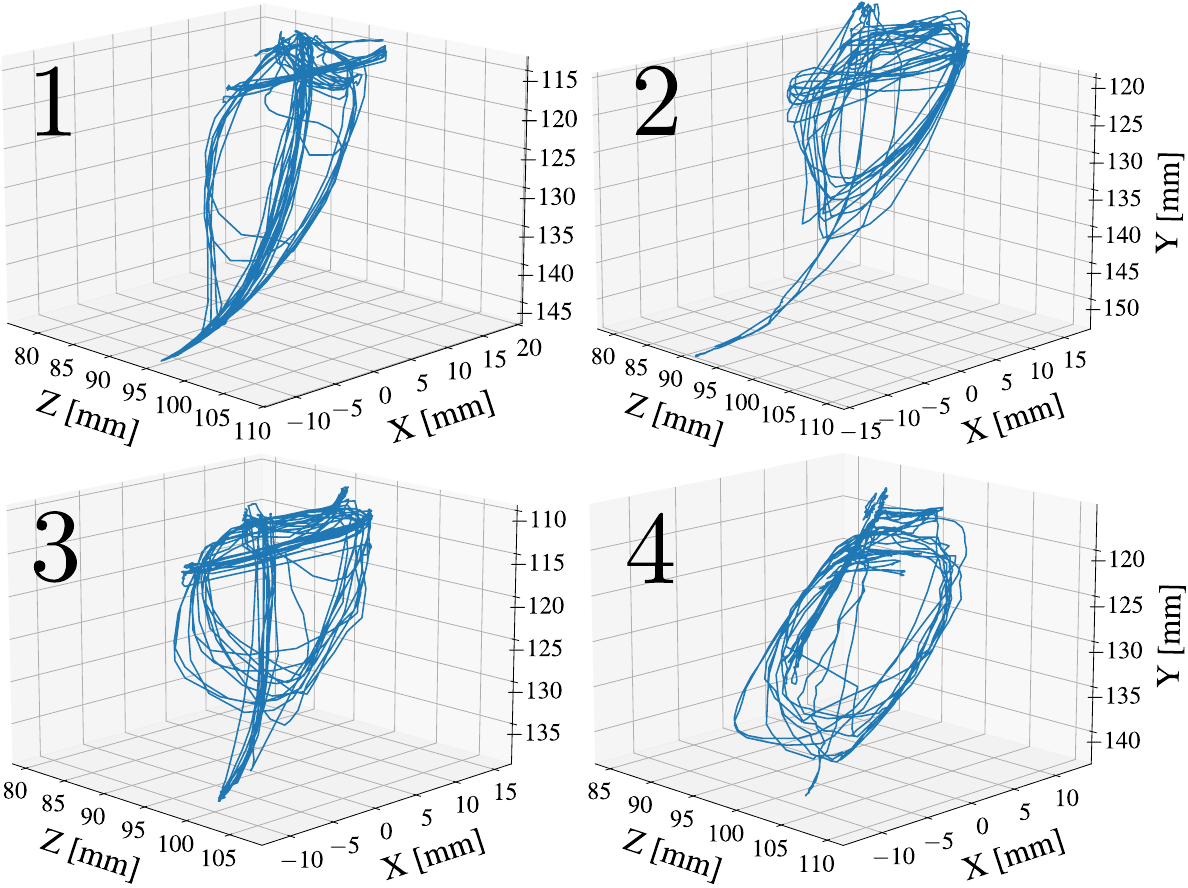}}
    \caption{The $3$D trajectories of all recorded jaw movements of the four participants, including opening and closing, protrusion and retrusion, lateral motions, and cyclic motions. The reference point on the jaw is the incisal point in the middle of the lower teeth.}
    \label{fig:jaw-tracking-data}
\end{figure}

\section{Experiments}

A preliminary study involving four participants (\SI{26.75 \pm 1.09}{\yearsold}) evaluated the jaw tracking system's functionality and performance. Ethics committee approval from the TU Darmstadt (approval number EK \num{3}/\num{2025}) was obtained, with participants providing written informed consent prior to study inclusion. Following MTA and CRA attachment and landmark recording, each participant performed a jaw movement series, including mouth opening and closing, left and right jaw movement, forward and backward motion, and cyclic motions. Data recording occurred at \SI{200}{Hz} using a \textit{Qualisys Oqus} (Qualisys AB, Göteborg, Sweden) OMoCap system with three cameras, processed according to the described pipeline. The same model was used for all participants for the jaw tracking data registration to a virtual jaw model, avoiding individual CT or MRI scans while accepting larger registration uncertainties due to anatomical variation between individuals.

\section{Results and Discussion}
The tracking system successfully recorded jaw movements for all participants, providing $3$D mandible trajectories (\autoref{fig:jaw-tracking-data}). The fourth participant had difficulties and limitations with some motions, explaining the slightly different-looking trajectory. The reference point for the trajectories was the incisal point of the lower teeth. 
The system's accuracy depends primarily on the OMoCap system performance, achieving \SI{0.6}{mm} average post-calibration accuracy in our setup, and tooth landmark registration accuracy. Minor additional errors potentially introduced by mouth attachment and glue elasticity can be mitigated through reinforced attachment and more rigid dental adhesive usage.

Precision and noise characteristic estimation involved raw signal filtering using a fourth-order Butterworth low-pass filter with a cutoff frequency of \SI{4.5 \pm 0.5}{Hz}, applied bidirectionally to prevent phase shift. We chose the cutoff frequency based on a \SI{10}{\decibel} drop in the estimated signal-to-noise ratio for each participant. The filtered signals represent the gross human motion, while we analyzed the residual signals to assess high-frequency noise content and approximate system precision. 
Mean and standard deviation calculations of the residual signals across recorded movements yielded \SI{182 \pm 47}{\mu\meter} and \SI{0.126 \pm 0.034}{\degree} across all participants. System precision thus proved worse than but comparable to commercial systems such as \textit{Modjaw} (\SI{9.7 \pm 1.76}{\mu\meter}) or \textit{Cyclops} (\SI{5.14}{\degree} - \SI{7.07}{\degree}) \cite{Nagy2023, Grande2024}. Due to the absence of reference data, we could not yet quantify the accuracy.

Regarding operational use, the system is designed for short-term evaluations and data acquisition in research rather than prolonged continuous monitoring. 
A maximum accuracy virtual jaw model registration requires individual patient CT or MRI scans for optimal anatomical landmark mapping. However, pre-existing generic jaw models might suffice for general kinematic analyses. Prior study validation demonstrated that data collected with the presented system, combined with jaw simulations, effectively reproduced recorded jaw movements and enabled jaw dynamics analysis \cite{Mueller2025a}.

Key system advantages include a modular design permitting straightforward customization and adaptation for diverse patients and clinical scenarios in an online or offline setting. Furthermore, the low-cost and open-source nature, encompassing both hardware and software, distinguishes it from many proprietary systems. This accessibility aims to foster rapid advancement in the nascent jaw exoskeleton field while providing a foundation for other researchers to adapt, extend, and improve the system.

\section{Conclusion and Outlook}  

We have presented a novel, customizable, low-cost, biocompatible, and open-source jaw movement tracking system utilizing OMoCap, usable offline or online depending on the OMoCap system's streaming capabilities \cite{Mueller2025c}. Identifying and digitizing six dental landmarks on the teeth obviates the need for custom marker-holding appliances such as braces or dental casts.
While the system architecture supports real-time streaming and processing of data, its performance has not been evaluated.
Future work includes comprehensive accuracy and performance validation through comparison with established tracking systems and reference trajectories \cite{Mostashiri2018, Farronato2021, Nagy2023}. This validation will involve phantom jaw models with precisely defined geometry and fiducial locations. Utilizing a spherical calibration tool tip may further enhance registration accuracy \cite{Abdi2018}.
We hope that this open-source jaw movement tracking system will support and enhance the design and development of novel jaw exoskeletons, enabling more effective rehabilitation strategies for TMDs.

\bibliographystyle{bibtex/IEEEtran}
\bibliography{bibtex/IEEEabrv,bibtex/sensors_2025}

\begin{thebibliography}{10}
\providecommand{\url}[1]{#1}
\csname url@samestyle\endcsname
\providecommand{\newblock}{\relax}
\providecommand{\bibinfo}[2]{#2}
\providecommand{\BIBentrySTDinterwordspacing}{\spaceskip=0pt\relax}
\providecommand{\BIBentryALTinterwordstretchfactor}{4}
\providecommand{\BIBentryALTinterwordspacing}{\spaceskip=\fontdimen2\font plus
\BIBentryALTinterwordstretchfactor\fontdimen3\font minus
  \fontdimen4\font\relax}
\providecommand{\BIBforeignlanguage}[2]{{%
\expandafter\ifx\csname l@#1\endcsname\relax
\typeout{** WARNING: IEEEtran.bst: No hyphenation pattern has been}%
\typeout{** loaded for the language `#1'. Using the pattern for}%
\typeout{** the default language instead.}%
\else
\language=\csname l@#1\endcsname
\fi
#2}}
\providecommand{\BIBdecl}{\relax}
\BIBdecl

\bibitem{Ahmad2016}
M.~Ahmad and E.~L. Schiffman, ``Temporomandibular joint disorders and orofacial
  pain,'' \emph{Dental Clinics of North America}, vol.~60, no.~1, pp. 105--124,
  Jan. 2016.

\bibitem{Gremillion2018}
H.~A. Gremillion and G.~D. Klasser, \emph{Temporomandibular disorders}.\hskip
  1em plus 0.5em minus 0.4em\relax Switzerland: Springer, 2018.

\bibitem{Durham2015}
J.~Durham, T.~R.~O. Newton-John, and J.~M. Zakrzewska, ``Temporomandibular
  disorders,'' \emph{BMJ}, vol. 350, no. mar12 9, p. h1154–h1154, Mar. 2015.

\bibitem{Tran2022}
C.~Tran, K.~Ghahreman, C.~Huppa, and J.~Gallagher, ``Management of
  temporomandibular disorders: a rapid review of systematic reviews and
  guidelines,'' \emph{International Journal of Oral and Maxillofacial Surgery},
  vol.~51, no.~9, p. 1211–1225, Sep. 2022.

\bibitem{Suppelt2025}
S.~Suppelt, R.~Chadda, N.~Schäfer, A.~A. Altmann, D.~Gerovac, D.~G.~E. Thiem,
  P.~Weigl, R.~Sader, and M.~Kupnik, ``Multiparameter measurement system for
  analyzing the temporomandibular joint complex,'' \emph{IEEE Sensors Journal},
  vol.~25, no.~4, pp. 7263--7275, Feb. 2025.

\bibitem{Shimada2019}
A.~Shimada, S.~Ishigaki, Y.~Matsuka, O.~Komiyama, T.~Torisu, Y.~Oono, H.~Sato,
  T.~Naganawa, A.~Mine, Y.~Yamazaki, K.~Okura, Y.~Sakuma, and K.~Sasaki,
  ``Effects of exercise therapy on painful temporomandibular disorders,''
  \emph{Journal of Oral Rehabilitation}, vol.~46, no.~5, pp. 475--481, Feb.
  2019.

\bibitem{Mueller2025b}
P.-O. Müller, R.~Sader, and O.~von Stryk, ``Exoskeletons for the
  rehabilitation of temporomandibular disorders: a comprehensive review,''
  \emph{Frontiers in Robotics and AI}, vol.~12, May 2025.

\bibitem{Scolaro2022}
A.~Scolaro, S.~Khijmatgar, P.~M. Rai, F.~Falsarone, F.~Alicchio, A.~Mosca,
  C.~Greco, M.~Del~Fabbro, and G.~M. Tartaglia, ``Efficacy of kinematic
  parameters for assessment of temporomandibular joint function and
  disfunction: A systematic review and meta-analysis,'' \emph{Bioengineering},
  vol.~9, no.~7, p. 269, Jun. 2022.

\bibitem{Nota2024}
A.~Nota, L.~Pittari, F.~M. Monticciolo, A.~C. Lannes, and S.~Tecco,
  ``Correlations between mandibular kinematics and electromyography during the
  masticatory cycle: An observational study by digital analysis,''
  \emph{Applied Sciences}, vol.~14, no.~21, p. 9996, Nov. 2024.

\bibitem{Morikawa2022}
K.~Morikawa, R.~Isogai, J.~Nonaka, Y.~Yoshida, S.~Haga, and K.~Maki, ``A new
  intraoral six-degrees-of-freedom jaw movement tracking method using magnetic
  fingerprints,'' \emph{Sensors}, vol.~22, no.~22, p. 8923, Nov. 2022.

\bibitem{Jucevicius2022}
M.~Jucevičius, R.~Ožiūnas, G.~Narvydas, and D.~Jegelevičius, ``Permanent
  magnet tracking method resistant to background magnetic field for assessing
  jaw movement in wearable devices,'' \emph{Sensors}, vol.~22, no.~3, p. 971,
  Jan. 2022.

\bibitem{Farronato2021}
M.~Farronato, G.~M. Tartaglia, C.~Maspero, L.~M. Gallo, and V.~Colombo, ``In
  vitro and in vivo assessment of a new workflow for the acquisition of
  mandibular kinematics based on portable tracking system with passive optical
  reflective markers,'' \emph{Applied Sciences}, vol.~11, no.~9, p. 3947, Apr.
  2021.

\bibitem{Nagy2023}
Z.~Nagy, A.~Mikolicz, and J.~Vag, ``In-vitro accuracy of a novel jaw-tracking
  technology,'' \emph{Journal of Dentistry}, vol. 138, p. 104730, Nov. 2023.

\bibitem{Grande2024}
F.~Grande, L.~Lepidi, F.~Tesini, A.~Acquadro, C.~Valenti, S.~Pagano, and
  S.~Catapano, ``Investigation of the precision of a novel jaw tracking system
  in recording mandibular movements: A preliminary clinical study,''
  \emph{Journal of Dentistry}, vol. 146, p. 105047, Jul. 2024.

\bibitem{Mueller2025c}
\BIBentryALTinterwordspacing
P.-O. Müller, S.~Suppelt, M.~Kupnik, and O.~von Stryk,
  ``{J}aw{T}racking{S}ystem ({JTS}): A customizable, low-cost, optical jaw
  tracking system,'' 2025. [Online]. Available:
  \url{https://github.com/paulotto/jaw_tracking_system}
\BIBentrySTDinterwordspacing

\bibitem{Kabsch1976}
W.~Kabsch, ``A solution for the best rotation to relate two sets of vectors,''
  \emph{Acta Crystallographica Section A}, vol.~32, no.~5, pp. 922--923, Sep.
  1976.

\bibitem{Mueller2025a}
P.-O. Müller and O.~von Stryk, ``The foundation for developing an exoskeleton
  for the rehabilitation of temporomandibular disorders,'' in \emph{2025 IEEE
  International Conference on Simulation, Modeling, and Programming for
  Autonomous Robots (SIMPAR)}.\hskip 1em plus 0.5em minus 0.4em\relax IEEE,
  Apr. 2025, pp. 1--6.

\bibitem{Mostashiri2018}
N.~Mostashiri, J.~S. Dhupia, A.~W. Verl, and W.~Xu, ``A novel spatial
  mandibular motion-capture system based on planar fiducial markers,''
  \emph{IEEE Sensors Journal}, vol.~18, no.~24, pp. 10\,096--10\,104, Dec.
  2018.

\bibitem{Abdi2018}
A.~H. Abdi, A.~G. Hannam, I.~K. Stavness, and S.~Fels, ``Minimizing fiducial
  localization error using sphere-based registration in jaw tracking,''
  \emph{Journal of Biomechanics}, vol.~68, pp. 120--125, Feb. 2018.

\end{thebibliography}

\end{document}